# Acoustic resonators above 100 GHz

Jack Kramer[1], Bryan T. Bosworth[2*], Lezli Matto[3*], Nicholas R. Jungwirth[2*], Omar Barrera[1], Florian Bergmann[2,4], Sinwoo Cho[1], Vakhtang Chulukhadze[1], Mark Goorsky[3], Nathan D. Orloff[2*], and Ruochen Lu[1]

[1]The University of Texas at Austin, Austin, 78712, TX, USA.

[2]National Institute of Standards and Technology, Boulder, 80305, CO, USA.

[3]University of California, Los Angeles, Los Angeles, 90095, CA, USA

[4]Department of Physics, University of Colorado, Libby Dr, Boulder, Colorado 80302, USA

*These authors contributed equally to the work.

Corresponding author: Jack Kramer, kramerj99@utexas.edu

## Abstract

Piezoelectric resonators are a common building block for signal processing because of their miniature size, low insertion loss, and high quality factor. As consumer electronics push to millimeter waves frequencies, designers must increase the operating frequency of the resonator. The current state-of-the-art approach to increase the operating frequency is to decrease the thickness of the piezoelectric film to shorten the acoustic wavelength or to use higher order modes. Unfortunately, maintaining high quality factors typically requires thicker piezoelectric layers. Thinner individual layers suffer from higher defect densities and increased relative losses from surface related damping, which degrade the electromechanical coupling and quality factor.

While acoustic high order modes can also increase operating frequency, the electromechanical coupling rapidly decreases with increasing mode number. Here, we overcome these limitations by utilizing a piezoelectric stack of three layers of lithium niobate with alternating crystallographic orientations to preferentially support higher order modes and thereby enhance the electromechanical coupling without degrading the quality factor. Our approach improves the figure of merit of millimeter-wave acoustic resonators by roughly an order of magnitude greater compared to state-of-the-art piezoelectric resonators above 60 GHz. This concept of alternating crystallographic orientations facilitates a path to develop millimeter wave resonators with high figures of merit, low insertion loss, and miniature footprints, enabling applications in millimeter wave signal processing and computing.

## Main

Piezoelectric materials couple electrical and mechanical fields[1]. By designing acoustic cavities from piezoelectric materials, engineers can create acoustic resonators that are orders of magnitude smaller than equivalent electromagnetic cavities and that take advantage of the low loss provided by acoustic waves[2–6]. These acoustic resonators are important building blocks for telecommunication filters[4,7], oscillators[8,9], acousto-optic modulators[10–12], and even quantum circuits[13,14]. Often the thickness defines the frequency of operation of the thin film acoustic cavity more than other geometry parameters[5,15,16] (Fig 1). Ranging from micrometers to hundreds of nanometers (Fig 1a), typical film thicknesses produce resonators that operate in the megahertz (MHz) to gigahertz (GHz) frequency ranges[17,18]. Thinner films can achieve higher operating frequencies, but surface losses drastically reduce the ability of the resonator to store energy in smaller volumes[19,20] , quantified by the resonator quality factor ($Q$) (Fig 1b). Along with decreasing $Q$, achieving thin films with high crystal quality poses a challenge. Defects degrade acoustic performance and result in poor electrical to mechanical transduction, quantified by the electromechanical coupling ($k^2$)[5]. One alternative to reducing film thicknesses is to use higher order acoustic modes (Fig 1a). Unfortunately, high order modes in an acoustic resonator reduce the electromechanical coupling since the sign of the local strain fails to match the crystal lattice orientation[21] (Fig. 1c), leading to charge cancellation. Consequently, acoustic resonators above 10 GHz tend to suffer from reduced figures of merit ($k^2Q$) (Fig 1b). Achieving a relatively high figure of merit in acoustic resonators in the tens and hundreds of GHz remains an open challenge. Solving this challenge could enable compact millimeter-wave communication front ends, high-speed integrated photonic modulation, and high-fidelity quantum phononic states.

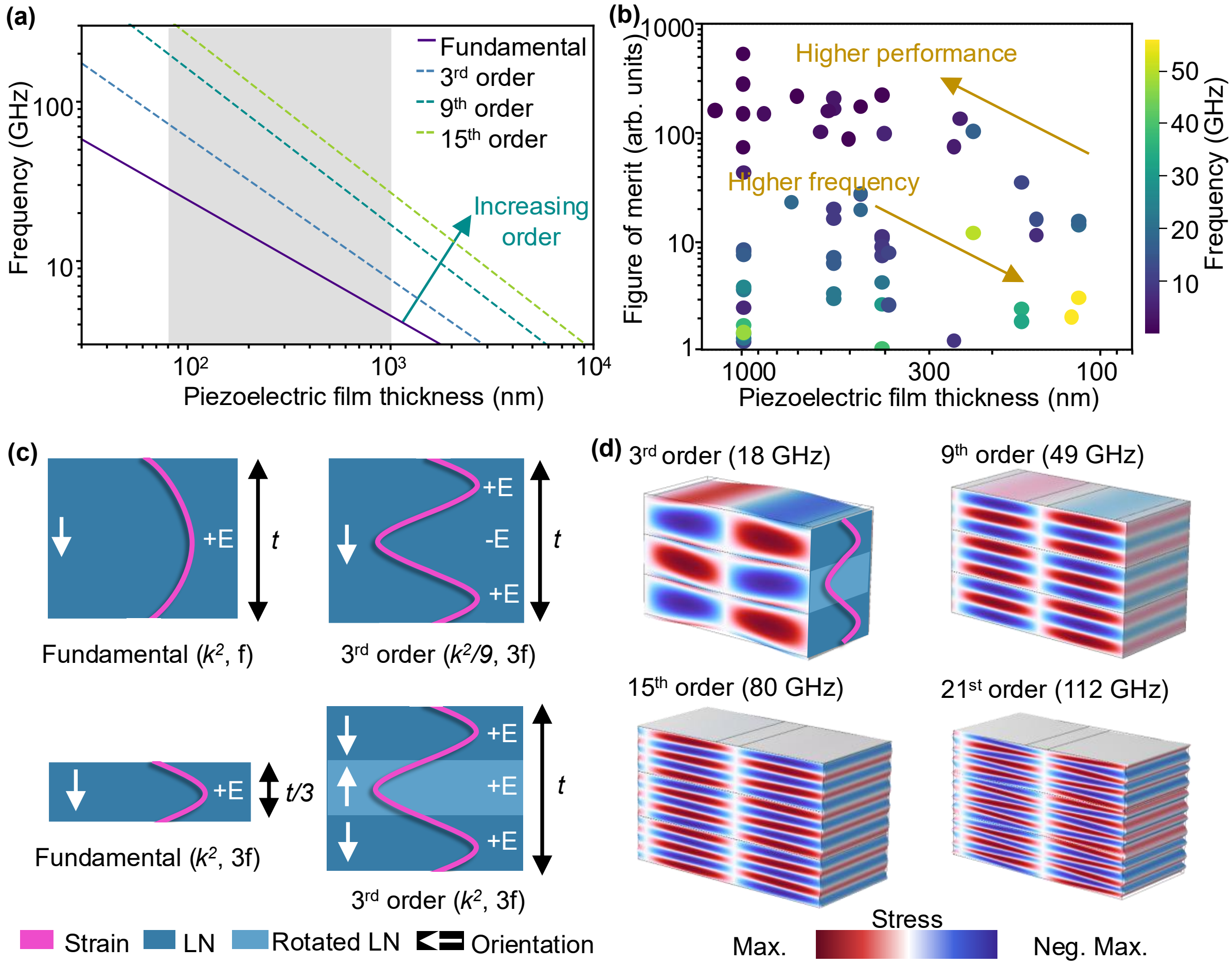
(a)
Frequency (GHz)
Fundamental
3rd order
9th order
15th order
Increasing order
Piezoelectric film thickness (nm)
(b)
Figure of merit (arb. units)
Higher performance
Higher frequency
Frequency (GHz)
Piezoelectric film thickness (nm)
(c)
+E
-E
t
t/3
Fundamental (k², f)
3rd order (k²/9, 3f)
Fundamental (k², 3f)
3rd order (k², 3f)
Strain
LN
Rotated LN
Orientation
(d) 3rd order (18 GHz)
9th order (49 GHz)
15th order (80 GHz)
21st order (112 GHz)
Stress
Max.
Neg. Max.

Figure 1| **(a)** Plot of acoustic mode resonant frequency as a function of piezoelectric film thickness. The limit of achievable high quality thin film piezoelectrics is highlighted in gray. **(b)** Survey of reported acoustic resonator figure of merit[15–17,21–45]. Figure of merit is defined as the quality factor, Q, times the electromechanical coupling, $k^2$. Thicker piezoelectric films yield higher figures of merit, while thinner films are required for high resonant frequency operation. **(c)** Illustration of impact of piezoelectric thickness, t, and crystal orientation on resonant frequency, f, and electromechanical coupling, $k^2$. Increasing mode order from the fundamental to the 3rd order triples the resulting resonant frequency, but at the cost of electromechanical coupling. Electromechanical coupling is reduced due to cancellation in the electric field, E, in the piezoelectric associated with the modal strain. The same frequency can also be achieved by reducing the thickness of the film, at the cost of quality factor and performance shown in **(b)**. However, electromechanical coupling and piezoelectric thickness are both maintained if the orientation of the piezoelectric is alternated such that the associated electric field adds constructively. **(d)** Simulated modal profiles for the 4 target modes in a three-layer lithium niobate. The modal profiles correspond to the 3rd, 9th, 15th and 21st thickness shear modes.

Transferred and bonded thin-film lithium niobate (LN) may offer a few solutions to help overcome some of the limitations in conventional acoustic resonators. First, the crystal quality of the transferred LN remains comparable to the bulk even at film thickness, which have relatively high defect densities when compared to deposited materials[46]. The transferred LN can also leverage low loss dielectric substrates with arbitrary crystal structure and orientation to reduce parasitic electromagnetic losses. Moreover, a sacrificial layer below the transferred LN can further confine acoustic energy within the cavity due to a high acoustic impedance mismatch between LN and air[22,47]. The last remaining challenge is the reduced electromechanical coupling of a high order mode resonator using a single-layer LN (Fig 1c). To overcome this, various researchers have begun examining using multiple layers of piezoelectric materials with alternating crystallographic orientation to address this challenge[23–27,45,48,49]. Alternating the crystallographic orientation of the piezoelectric material could match the local electric field profile to the strain among layers, resulting in a constructive contribution to the electromechanical coupling. When realized through a thin film transfer process, independent selection of the independent layer crystallographic orientation could be made without needing to consider the substrate crystalline structure. Through this approach, transferred LN in multiple layers with tailored thicknesses and crystallographic orientation, has shown potential to push the technology envelope embodied by the tradeoff between operating frequency and the acoustic figure of merit—electromechanical coupling ($k^2$) times quality factor (Q). In this work, build upon preliminary results at sub-50 GHz to enable acoustic resonators designed for much higher frequencies[23,24]. This expands upon the prior results by implementing a miniaturized high quality 3- layer stack, improved acoustic design, and utilizing precise on-chip measurement techniques. This allows us to achieve a large enough signal-to-noise ratio to directly measure acoustic resonances to 120 GHz above 100 GHz (Fig. 1d).

Here, we test an approach to design acoustic resonators above 100 GHz by transferring multiple LN layers into a three-layer stack with alternating crystallographic orientations. After transferring, we validate the crystal quality of each layer in the multilayer stack. We discuss the codesign and co-fabrication of acoustic resonators with microwave calibration kits, enabling the acoustic response to be de-embedded from electromagnetic effects. We perform S-parameter microwave characterization of the acoustic resonators with on-wafer techniques to 220 GHz. From the measured data, we observe electrically excited acoustic modes from 16 to 120 GHz, enabled by the multilayer LN. Beyond 60 GHz, the extracted figures of merit exceed the state-of-the-art. With further optimization and integration, the multilayer LN stack platform advances toward the viable deployment of miniature acoustic devices above 100 GHz.

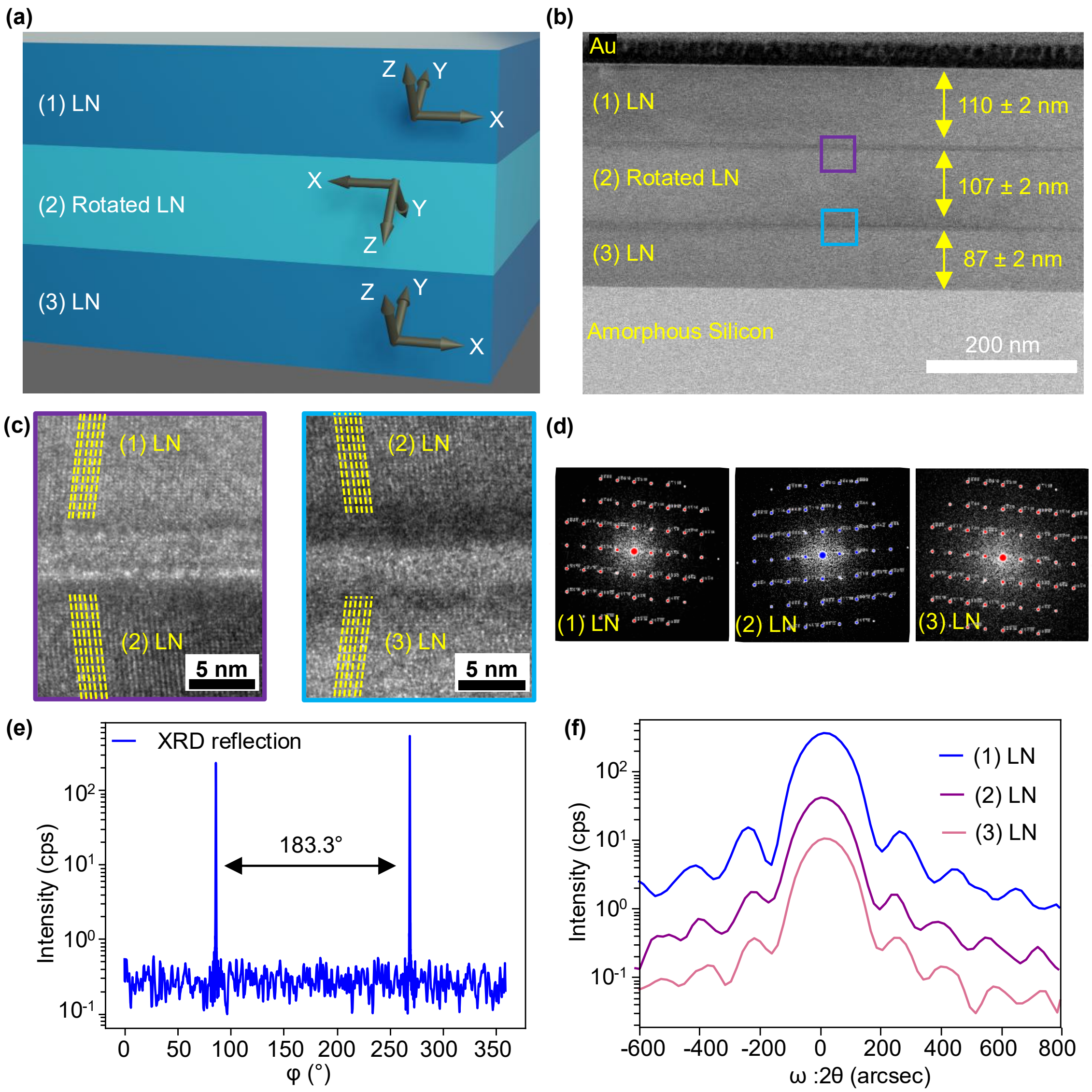

Figure 2| **(a)** Illustration of the multilayer lithium niobate stack with alternating crystal orientations. The crystal orientation of each layer is indicated by the illustrated axis. **(b)** Bright field scanning transmission electron microscope image of the multilayer lithium niobate stack. The purple and blue outlined regions correspond to the high-resolution scanning transmission electron shown in panel **(c)**. **(c)** High resolution STEM images of layer interfaces, showing alternating crystallographic orientations. An amorphous interfacial layer from the bonding process is also observed. **(d)** Superimposed observed and simulated electron diffraction patterns of the individual

lithium niobate layers. The red and blue dots correspond to the simulated diffraction patterns, while the white dots are the observed diffraction patterns. **(e)** ϕ x-ray diffraction scan measurement of the multilayer lithium niobate stack. Two peaks are observed, corresponding to the two desired crystal orientations, offset by 183.3 degrees. The two peaks confirm the in-plane orientation of the layers is close to the desired 180-degree offset. The reduced first peak is due to the existence of two closely spaced peaks, associated with layers (1) and (3) of the lithium niobate. More information is supplied in supplementary information. **(f)** ω:2θ x-ray diffraction scan of the multilayer stack. Each of the three layers displays strong fringes associated with preserved high crystal quality.

To enable acoustic resonances above 100 GHz, we chose three layers of alternating LN to target the 3$^{rd}$, 9$^{th}$, 15$^{th}$ and 21$^{st}$ order thickness shear modes (Fig. 2a, b). In a three-layer LN stack, the superlattice preferentially supports the 3$^{rd}$ order mode with the latter modes being the higher order modes that also provide the highest electromechanical coupling. To optimize for the 3$^{rd}$ shear mode, we selected 128Y cut LN layers for the maximum value of the shear $e_{15}$ component of the electromechanical coupling matrix. The ideal configuration of the multilayer LN consists of alternating upwards and downwards facing 128Y cut LN layers (Fig. 2a) that have been aligned such that their crystallographic X-axis are rotated 180° opposite one another. Proper alignment of the crystallographic orientations is important for mitigating the excitation of additional acoustic modes and for maximizing the electromechanical coupling of the targeted shear modes. To achieve the ideal electromechanical coupling for the shape of the 3$^{rd}$ order shear mode, alternating layers were nearly equal in film thicknesses relative to each other (Fig. 2b). This thickness requirement made it challenging to realize practical multilayer transferred LN[19]. It is possible that additional layers could improve the figure of merit of higher order modes; however, we expect some degradation in the figure of merit as the thickness of the transferred LN approaches the effective thickness of the interfacial layer (Fig. 2c).

Our idea assumes that the process of micromachining and transferring each LN layer does not impact crystal quality and the interfacial material properties are negligible. The bright field scanning transmission electron microscopy (STEM) showed three distinct LN layers and confirmed their respective thicknesses (Fig. 2b). While we observed slight differences in layer thicknesses, each layer remained within the regime where we expected enhanced electromechanical coupling. At the interfaces of the LN layers (Fig 2c), we observed the alternating crystallographic orientation of the LN. Both electron diffraction patterns and Fourier transform (SFig. 1) of the lattice images from the alternating layers (FFT) produced reciprocal space diffraction patterns for each layer and showed the expected upwards and downwards facing 128Y cut LN orientations (Fig. 2d). To further ensure the correct crystallographic orientations and crystalline properties, we performed various high-resolution X-ray diffraction (HR-XRD) scans about different axes (Fig. 2e). An HR-XRD ϕ scan validated the in-plane rotation of the LN layers, which displays a 183° rotation of the intermediate layer X-axis orientation relative to the top and bottom layers.

We observed an in-plane misalignment of less than 1° for the top and bottom layers relative to each other. Using an HR-XRD ω:2Θ scan, we extracted film thicknesses through the spatial separation of the thin film interference fringes for each of the layers (Fig. 2f). The observed fringe spacing corresponds to thicknesses of 98 nm, 98 nm, and 87 nm for LN layers 1, 2, and 3 respectively, using the nomenclature in Figure 2. The layer thickness differences are very small considering the bonding, lapping and chemical mechanical polishing (CMP) steps used to create each layer in succession across a 4-inch wafer. The measurement location of the BF-STEM and HR-XRD differed slightly from each other, resulting in slight variation in thickness between the two methods. HR-XRD rocking curve measurements verified that the transferred LN crystal retained its initial quality after the thinning process. The various layers had full width at half maximum (FWHM) peaks ranging from 45 to 260 arcseconds, confirming the high overall crystalline quality of these layers (SFig. 2). These small FWHM values show that the LN quality is preserved through the formation of the multilayer LN stack, potentially enabling high performance acoustic resonators.

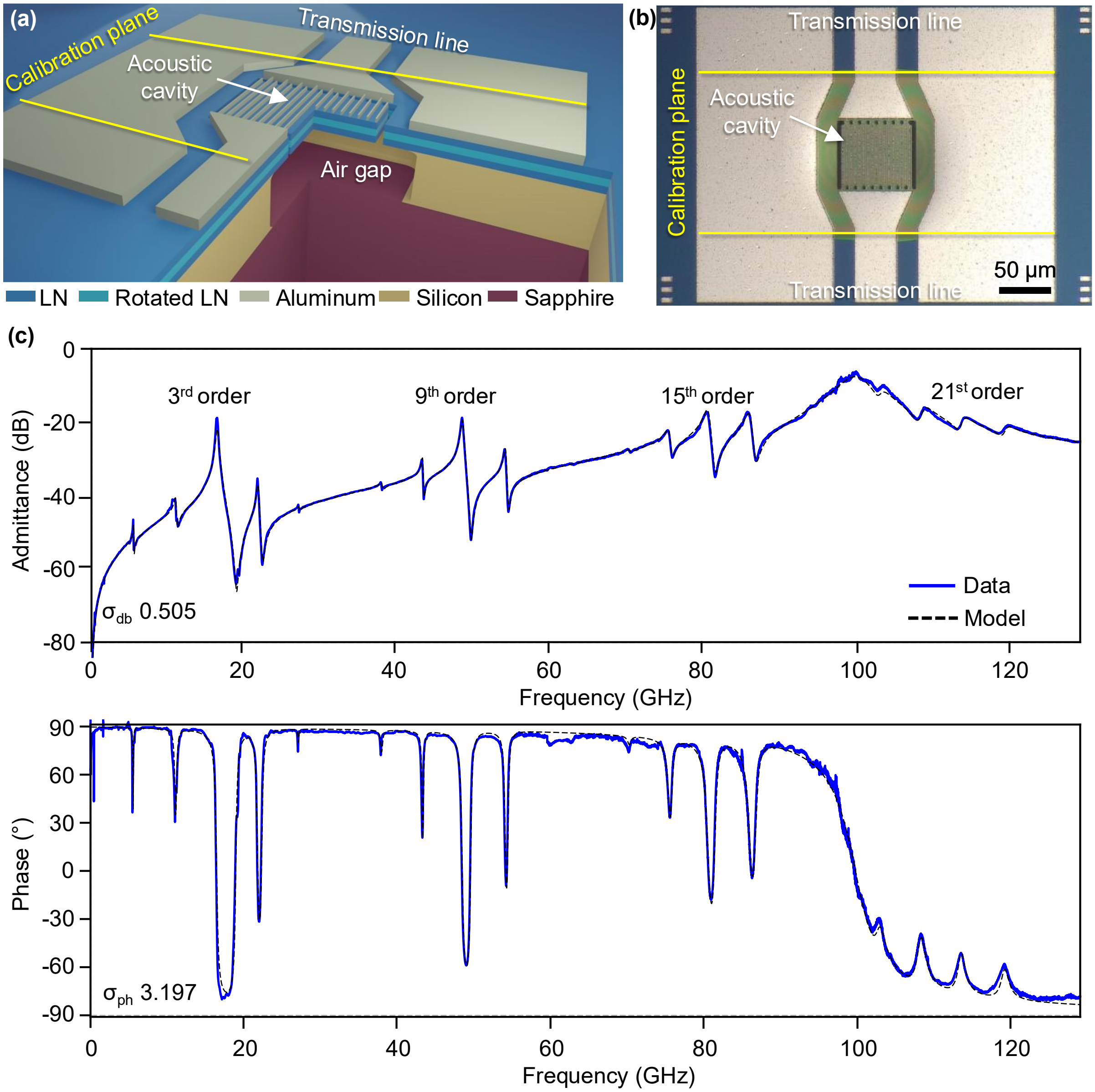
(a)
Calibration plane
Transmission line
Acoustic cavity
Air gap
LN
Rotated LN
Aluminum
Silicon
Sapphire
(b)
Transmission line
Calibration plane
Acoustic cavity
50 μm
Transmission line
(c)
3rd order
9th order
15th order
21st order
Admittance (dB)
Data
Model
σdb 0.505
Frequency (GHz)
Phase (°)
σph 3.197
Frequency (GHz)

Figure 3| **(a)** Illustration of the acoustic resonator topology. An acoustic cavity is formed, isolated from the rest of the multilayer stack and the substrate with air gaps. The cavity is actuated by aluminum electrodes that span across the surface. A microwave transmission line allows for on-chip measurements using a vector network analyzer. An on-chip calibration kit with matching dimensions allows for the translation of the calibration plane to the tapered transmission line region. Details are provided in the supplementary information. **(b)** Optical micrograph of the fabricated acoustic cavity. The region that has been isolated from the substrate is indicated by the change in color, from blue to yellow, of the film. **(c)** Broadband admittance measurement of the acoustic resonator. Resonant modes are observed up to 120 GHz. The targeted 3rd, 9th, 15th and 21st order modes show the strongest peaks, associated with the large electromechanical coupling. Adjacent modes are also observed neighboring the target modes. These are associated with thickness variance between individual lithium niobate layers, with additional information provided in the supplementary. The model uses a 17-branch modified Butterworth-van Dyke circuit and is used for device performance extraction. Details are provided in the supplementary.

Today, on-wafer metrology above 100 GHz is not routine because the required measurement equipment is uncommon. On-wafer calibration kits should be tailored to the application[50] (SFig. 3). Above 100 GHz, one must also consider the resonator (SFig. 4b,c) along with the de-embedding test structures (SFig. 4d-g) during design to accurately characterize the response of the acoustic cavity (Fig. 3). Isolating the acoustic cavity both electrically and mechanically from the environment helped maximize the quality factors and minimize parasitic modes. Our acoustic resonator consisted of an acoustic cavity and an aluminum microwave transmission line (Fig. 3a) electrically designed to minimize radiation and surface waves[51]. An air gap (Fig 3a) mechanically decoupled the acoustic cavity from the substrate. The interdigitated aluminum electrodes electrically excited the acoustic waves in the cavity. We selected aluminum for its comparatively low resistivity, low density, and its compatibility with the sacrificial layer etch process (Supplemental, SFig. 5). 50 µm pitch ground-signal-ground probes electrically contacted each resonator, enabling measurement with an ultrabroadband 220 GHz vector network analyzer (VNA) (Supplemental, SFig. 6).

The ultrabroadband measurement used a two-tier calibration approach from 10 MHz to 220 GHz. The first tier was an off-chip calibration on sapphire with a custom on-wafer calibration kit that had nominally identical cross-sectional dimensions as the transmission lines on our resonators. The line lengths of the first-tier calibration minimized the uncertainty in the calibration over the bandwidth[50]. This first-tier calibration translated the measurement reference plane to the probe tips and set the reference impedance to 50 $\Omega$. Next, we performed a second-tier calibration with the on-wafer calibration kit, which was also designed to minimize the uncertainty in the calibration. This second-tier calibration translated the reference plane to the resonator (Fig. 3b). The second-tier calibration utilized the same well-defined transmission as the device and shifted the reference plane. The second-tier calibration included the effects of the probing pad region[52], differentiating it from techniques such as short-open de-embedding which remove these effects. We cascaded the first and second tier calibrations to correct the measurements of the acoustic resonators. The measured broadband response showed resonant peaks associated with acoustic modes from 10 GHz to 120 GHz (Fig 3c).

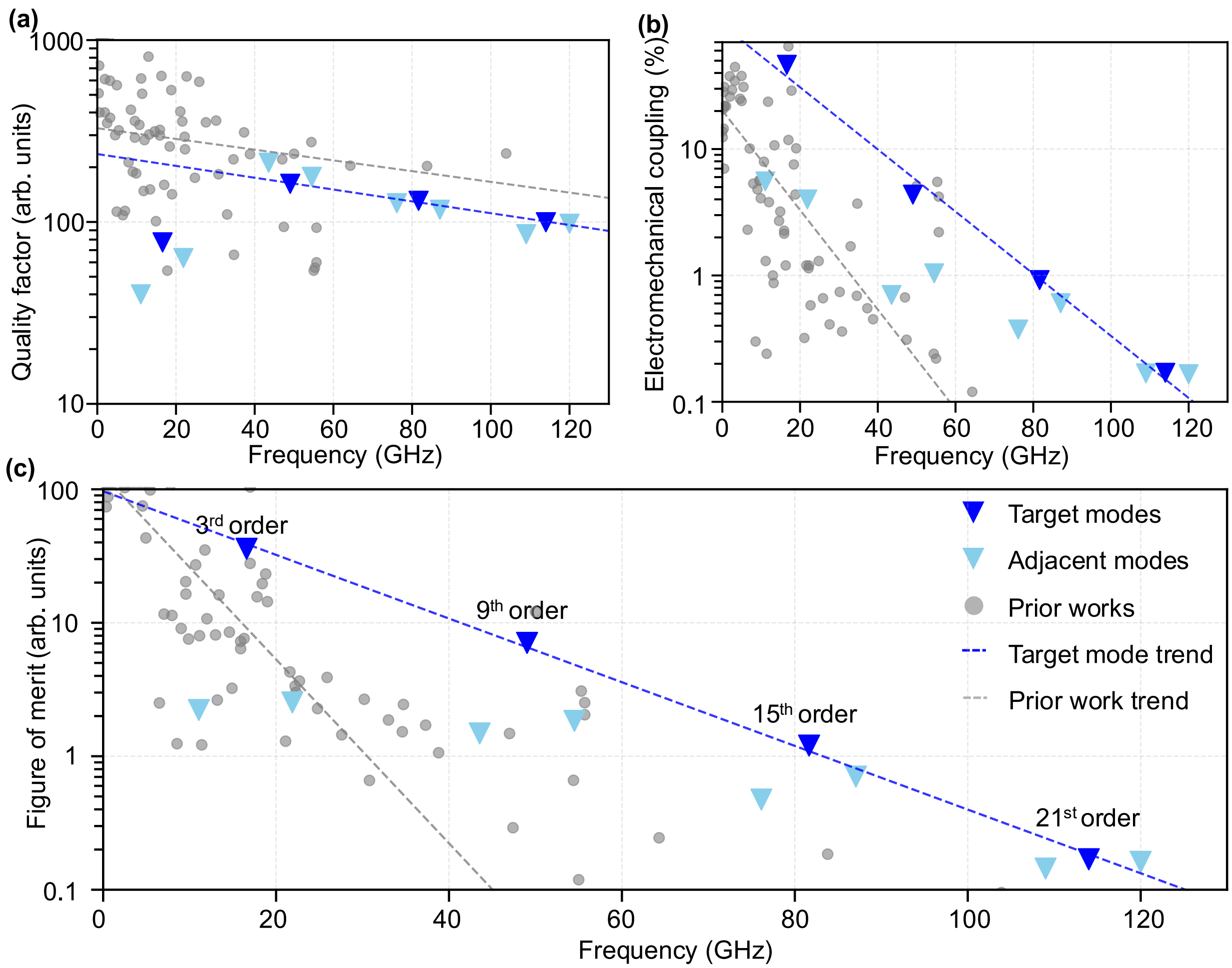

Figure 4|. (a), (b), (c) Plot of quality factor, electromechanical coupling, and figure of merit respectively, for the target modes and adjacent modes. An additional survey of prior works[15–17,21–45] is also included for reference. A robust linear fitting, which neglects statistical outliers, is applied to target modes and prior work to provide the data trend. Higher electromechanical couplings are achieved through the multilayer lithium niobate platform. Quality factors are comparable, on the order 200 and degrade to 100 for the modes beyond 100 GHz. The combination allows for a higher resultant figure of merit, particularly for the target modes towards and beyond 100 GHz. While this figure of merit is lower than sub-10 GHz, the achieved signal-to-noise ratio enables for direct measurement.

The strongest coupling acoustic resonance near 16 GHz corresponds to the enhanced 3$^{rd}$ order shear mode, with the next target modes being the peaks at roughly 50, 80, and 115 GHz. Adjacent to the target modes, there are acoustic resonances corresponding to the n +/- 1 order shear modes. These adjacent modes theoretically receive significantly lower electromechanical coupling than the target modes, but the practical value fluctuates with the relative layer thicknesses within the multilayer LN stack (SFig. 7, SFig. 8). The general character of the curve (Fig. 3c) is consistent with an electromagnetic resonance at roughly 100 GHz. A series inductance in the tapered transmission line and the static capacitance of the transducer electrodes were the dominant causes of this electromagnetic effect (SFig. 9). We confirmed that the frequency of this electromagnetic resonance depends directly on the acoustic resonator geometry (SFig. 10, STable 2) and the corresponding impact on these two values. As an example, we noted that increased electrode spacing leads to higher frequency electromagnetic resonances due to the reduced static capacitance (SFig. 11). Using the measured device data, we construct a multibranch modified Butterworth-Van Dyke circuit model[53] of the acoustic resonator (SFig. 9) that extracts (STable 1) the quality factor ($Q$) and electromechanical coupling ($k^2$) of each represented resonance. The circuit fitting extracted the performance metrics instead of employing 3dB Q or other methods due to interplay of mechanical and electrical effects that lead to distorted raw extraction metrics.

The acoustic resonators on transferred multilayer LN demonstrated comparable quality factors (Fig. 4a) with enhanced electromechanical coupling (Fig. 4b) relative to the state-of-the-art (grey dashed line, Fig. 4a,b). The extracted quality factors for the given modes remained comparable to the state-of-the-art (Fig. 4b). The measured enhanced coupling (Fig. 4b) upholds the hypothesis that the multilayer LN stack preferentially supports acoustic modes where the sign of the local strain matches the crystal lattice orientation in each layer. The resultant figure of merit (blue dashed line, Fig. 4c) represents a significant improvement when compared to the state-of-the-art (grey dashed line, Fig. 4c). The figures of merit follows the trend set by the decreased electromechanical coupling of the acoustic higher-order modes. While the figures of merit are low compared to sub-10 GHz works, these are sufficiently high to allow for direct measurement of the acoustic modes, beyond the noise limit of the network analyzer.

Further work is now possible towards determining what limits in resonator figure of merit are possible for acoustic technology beyond 100 GHz. Achieving yet higher frequency resonators faces the challenge of further improving the $Q$ to bring improve the signal to noise ratio of the resonance. The current trend shows a $Q$ that is nearly constant over frequency, which warrants further investigation. The perceived limit observed in this work suggests that surface related damping mechanisms are likely no longer the limiting factor in this frequency range, motivating future work on the subject.

These results underscore the achievable improvement possible in millimeter wave acoustics through the preferential enhancement of high order modes with multilayer lithium niobate with alternating crystallographic orientation. This approach achieves higher electromechanical coupling to enable higher figures of merit and the highest frequency demonstration of electrically excited acoustic resonators. The figure of merit improvement achieved creates an opportunity for multilayer lithium niobate resonators and filters above 100 GHz. For example, our 49 GHz resonance with high figure of merit of 7 could enable a filter with an insertion loss of 2.4 dB and a fractional bandwidth 4.6%, reducing the receiver amplifier power by over half (SFig. 12). Especially for spaceborne applications, we estimate this filter would weigh only 750 µg, compared to approximately 40g for space-qualified cavity filters. Additionally, efficient high frequency acoustic transduction could allow on-chip photonic mixers and modulators, where thin-film lithium niobate is already broadly explored in a single-layer configuration[54]. The high frequencies achieved here could also allow for higher temperature hybrid quantum circuits. For example, from Bose-Einstein statistics, we estimate that our 114 GHz resonance has less than 1% thermal occupation probability at 1.2K, while a 1 GHz resonance would require a temperature below 10.5mK to achieve this thermal occupation[55]. The multilayer lithium niobate topology could be further expanded to include more layers or layers of different thicknesses, orientations, or even different materials. More generally, the multilayer transfer approach could be advantageous for enabling thin film systems with substantially different properties from a single layer counterpart. The multilayer transfer approach opens the doors for expanding the functionality of thin film technology.

The project was funded by Defense Advanced Research Projects Agency (DARPA) Department of Science Office COmpact Front-end Filters at the ElEment-level (COFFEE) project under grant number HR0011-22-2-003. The authors would like to thank Dr. Ben Griffin, Dr. Todd Bauer, and Dr. Zachary Fishman for their helpful discussions. The authors would like to acknowledge the helpful discussions with Dr. Keji Lai, Dr. David Burghoff, Dr. Ari Feldman, and Dr. Jeffrey Jargon during the preparation of the manuscript.

The data that support the findings of this study are available from the corresponding author upon reasonable request.

## Supplementary information

In the supplementary information, we include description of the methods used for testing the devices, the fabrication process used to realize the designs, a detailed description of the calibration process, analysis of the effect of layer variation, the impact of electromagnetic resonance, and how the current designs could realize high frequency filters. Additional figures are provided with details of the material analysis as well.